\documentclass{article}
 \input{epsf}
\usepackage{amsmath}
\usepackage{epsfig}
\usepackage{amssymb,amsfonts}

\numberwithin{equation}{section}

\def\p{\partial}

\def\pbar{\bar{\p}}
\def\zbar{\bar z}

\def\vbar{\overline v}
\def\Tr{{{\rm Tr~ }}}
\def\tr{{\rm tr\ }}
\def\Re{{\rm Re\hskip0.1em}}
\def\Im{{\rm Im\hskip0.1em}}
\def\irt2{\frac{1}{\sqrt{2}}}

\def\b{\beta}
\def\a{\alpha}

\font\cmss=cmss10
\font\cmsss=cmss10 at 7pt
\def\IL{\relax{\rm I\kern-.18em L}}
\def\IH{\relax{\rm I\kern-.18em H}}
\def\IR{\relax{\rm I\kern-.18em R}}
\def\inbar{\vrule height1.5ex width.4pt depth0pt}
\def\IC{\relax\hbox{$\inbar\kern-.3em{\rm C}$}}
\def\rlx{\relax\leavevmode}
\def\ZZ{\rlx\leavevmode\ifmmode\mathchoice{\hbox{\cmss Z\kern-.4em Z}}
 {\hbox{\cmss Z\kern-.4em Z}}{\lower.9pt\hbox{\cmsss Z\kern-.36em Z}}
 {\lower1.2pt\hbox{\cmsss Z\kern-.36em Z}}\else{\cmss Z\kern-.4em
 Z}\fi}
\def\IZ{\relax\ifmmode\mathchoice
{\hbox{\cmss Z\kern-.4em Z}}{\hbox{\cmss Z\kern-.4em Z}}
{\lower.9pt\hbox{\cmsss Z\kern-.4em Z}}
{\lower1.2pt\hbox{\cmsss Z\kern-.4em Z}}\else{\cmss Z\kern-.4em
Z}\fi}
\def\be{\begin{equation}}
\def\ee{\end{equation}}
\def\ba{\begin{align}}
\def\ea{\end{align}}
\def\beq{\begin{eqnarray}}
\def\eeq{\end{eqnarray}}
\begin{document}

\title{\bf A Twisted Non-compact Elliptic Genus}

\author{
Sujay K. Ashok$^{a}$ and Jan Troost$^{b}$}

\date{}

\maketitle

\begin{center}

\emph{$^{a}$Institute of Mathematical Sciences\\
  C.I.T Campus, Taramani\\
  Chennai, India 600113\\ }

 \emph{\\${}^{b}$ Laboratoire de Physique Th\'eorique}\footnote{Unit\'e Mixte du CNRS et
    de l'Ecole Normale Sup\'erieure associ\'ee \`a l'universit\'e Pierre et
    Marie Curie 6, UMR
    8549.
}
\\
\emph{ Ecole Normale Sup\'erieure  \\
24 rue Lhomond \\ F--75231 Paris Cedex 05, France}
\end{center}

\begin{abstract}
  We give a detailed path integral derivation of the elliptic genus of
  a supersymmetric coset conformal field theory, further twisted by a
  global $U(1)$ symmetry.  It gives rise to a Jacobi form in three
  variables, which is the modular completion of a mock modular form.
  The derivation provides a physical interpretation to the
  non-holomorphic part as arising from a difference in spectral
  densities for the continuous part of the right-moving bosonic and
  fermionic spectrum. The spectral asymmetry can also be read off
  directly from the reflection amplitudes of the theory. By performing
  an orbifold, we show how our twisted elliptic genus generalizes an
  existing example.
\end{abstract}

\newpage

\tableofcontents

\section{Introduction}
The elliptic genus of two-dimensional conformal field theories
with two supersymmetries provides useful information about their
spectrum \cite{Schellekens:1986yi}\cite{Witten:1986bf}. It is
protected by supersymmetry, and can be computed in a convenient
corner of moduli space. To acquire more information, the elliptic
genus can be further twisted by global symmetries of the model.
Elliptic genera have applications to anomaly calculations in
string theory, to algebraic geometry, to the study of renormalization group
flows, as well as to black hole entropy counting (see e.g.
\cite{Witten:1993jg}\cite{Eguchi:2010xk}\cite{Gaiotto:2008cd}\cite{Manschot:2009ia}\cite{DMZ}).

Elliptic genera of sigma-models with non-compact target spaces
exhibit further subtleties compared to their compact counterparts.
They give rise to Jacobi forms which contain non-holomorphic
contributions necessary to render mock modular holomorphic
expressions modular \cite{Zwegers}\cite{Zagier}.  In physical applications, these
contributions are often postulated under the hypothesis that a
certain duality group, the modular group, is a symmetry of the
model.

In \cite{Troost:2010ud} a derivation of the
non-holomorphic part of the elliptic genus of an orbifold of
 the non-compact coset
conformal field theory $SL(2,\mathbb{R})/U(1)$ was given.  The path
integral result was modular, and contained both the mock modular
contributions as well as its non-holomorphic completion. The elliptic
genus functioned as a modular covariant infrared cut-off to the bulk
partition function, exhibiting localized supersymmetric contributions,
as well as contributions from long multiplets. The latter were made
possible through the cancellation of a volume divergence with a
fermion zero-mode.

In this paper, we will derive the path integral result in more
detail, and directly in the axial coset conformal field theory in
section \ref{path}.  We further twist the elliptic genus with a
global $U(1)$ symmetry of the model. In section
\ref{contributions} we lay the link to the theory of mock modular
forms.  We moreover provide in section \ref{remarks} an
independent derivation of the measure of integration in the
non-holomorphic remainder term via the evaluation of the spectral
asymmetry as the derivative of the difference in phase shifts for
right-moving bosons and fermions. We also recuperate and
generalize the results of \cite{Troost:2010ud} by performing a
$\mathbb{Z}_k$ orbifold on the axial coset, where
$\mathbb{Z}_k$ is a subgroup of the $U(1)_R$ symmetry of the model.

In  recent independent work \cite{Eguchi:2010cb}, the results of
\cite{Troost:2010ud} were also extended towards the axial coset,
and further to models with fractional levels as well as to
orbifold sectors. One main point of \cite{Eguchi:2010cb} is an
analysis of the ordinary bulk partition function and the modular completion
of the discrete contribution to the partition function. We
concentrate on the extension of the results of
\cite{Troost:2010ud} to include a new global $U(1)$ twist.

\section{The path integral}
\label{path} In this section, we compute the path integral of the
supersymmetric coset conformal field theory $SL(2,\mathbb{R})/U(1)$ at integer level
$k$, twisted by left
R-charge as well as a global $U(1)$ symmetry. This is heavily based on the calculation
of the bosonic bulk partition function \cite{Hanany:2002ev} as well as the treatment
of the supersymmetric model in \cite{Eguchi:2004yi}, though there are differences in 
the details. See
\cite{Israel:2004ir} for a construction of the supersymmetric
partition function by the technique of deformation. A generalization of our analysis
at least for fractional levels $k$ exists.

\subsection{The Supersymmetric Axially Gauged WZW model}

\subsubsection{Bosons and Fermions}

The action $S_b$ for the bosonic axially gauged WZW model can be written
as $S_b = \kappa \, I(g,A)$ where we define \cite{BarsSfetsos}:
\be\label{SAinitial} I(g,A) = I(g) +\frac{1}{2\pi}\int d^2z\ \Tr
\left[2A_z\pbar g g^{-1} - 2\tilde
  A_{\zbar} g^{-1}\p g + 2g^{-1}A_z g\tilde A_{\zbar} - A_{z}
  A_{\zbar}-\tilde A_{z}\tilde A_{\zbar} \right], \ee
and $\kappa$ is a constant prefactor.
Here $I(g)$ refers to the action of the WZW model with the group valued map $g$:
\be I(g) = \frac{1}{2\pi}\int d^2 z\ \Tr(g^{-1}\p g g^{-1}\pbar g)
+ \frac{i}{12\pi}\int d^3z\ \Tr ((g^{-1}dg)^3)\,. \ee The gauge
fields are defined such that
\be\label{gaugefield} A_z = T_a\, A_z^a\qquad A_{\zbar} = T^a\,
A_{\zbar}^a\quad \tilde A_z = \tilde T_a\, A_z^a\qquad \tilde
A_{\zbar} = \tilde T_a\, A_z^a.  \ee
The notations $T^a$ and $\tilde{T}^a$ refer to the generators of the Lie algebra
whose exponentiation is the gauged subgroup $H\subset G$. There are only two independent gauge field components: 
the gauge field
$\tilde A$ differs only in the distinct embedding of the generators $\tilde{T}^a$
into the gauged subgroup. We introduce the subgroup-valued fields $h, \tilde{h},\bar{h},
\tilde{\bar{h}}$ such that:
\begin{align}\label{gaugefieldsh}
A_z            &=\p h h^{-1}\quad \text{and}\quad A_{\zbar}= \pbar
\bar h\ \bar h^{-1}\cr \tilde A_{z}&=\p\tilde h \tilde h^{-1}\quad
\text{and}\quad  \tilde{A}_{\zbar} = \pbar \tilde{\bar h}\
\tilde{\bar h}^{-1} \,,
\end{align}
and obtain the action in the form
\begin{multline}\label{s1+s2+s3+s4}
I(g,A)-I(g) = \frac{1}{2\pi}\int d^2z\ \tr\left[2\p h h^{-1} \pbar g g^{-1} -2\pbar \tilde{\bar h} \tilde{\bar h}^{-1} g^{-1}\p g \right.\cr
\left. \hspace{1in}+2\p h h^{-1} g \pbar \tilde{\bar h} \tilde{\bar h}^{-1}g^{-1} - \pbar \bar h \bar h^{-1}\p h h^{-1} - \pbar \tilde{\bar h} \tilde{\bar h}^{-1}\p \tilde{h} \tilde{h}^{-1}\right]\,.
\end{multline}
The action has the following gauge symmetry:
\begin{align}
g&\rightarrow m\, g\,  \tilde{m}^{-1} \cr h&\rightarrow m\, h
\qquad \bar h\rightarrow m\, \bar h \cr \tilde h&\rightarrow
\tilde m\, \tilde h \qquad \tilde{\bar h}\rightarrow \tilde m\,
\tilde{\bar h} ,
\end{align}
where $m$ takes values in the gauge group $H$ and $\tilde{m}$ is
defined in terms of $m$ as above.
We consider the gauged subgroup $H$ to be $U(1)$ and pick the
anomaly free
axial gauging $\tilde T = -T$. Substituting this into the
expression for the gauge field we find that we have the relations:
\be
\tilde h = h^{-1} \quad\text{and}\quad \tilde{\bar h} = \bar h^{-1} \,.
\ee
We will study the theory in which the group element $g$ after
analytic continuation takes values in Euclidean $AdS_3$. This is
the hyperbolic three-plane corresponding to the space of $2\times
2$ hermitian matrices $g$ with unit determinant. We will use the
following parameterization of these matrices:
\be\label{gads3}
g= \left(\begin{array}{cc}
e^{-\phi} & v \cr
\vbar & e^{\phi}(1+v\bar v)
\end{array}\right)\,.
\ee
Substituting this into the WZW action, one can write down the world sheet action for the bosonic fields:
\begin{align}
I(g) &= \frac{\kappa}{\pi}\int d^2z\, \bigg(\p\phi \pbar\phi+(\p \vbar + \vbar\, \p\phi)(\pbar v + v\, \pbar \phi) \bigg) \,.
\end{align}
The axially gauged action, with gauge field $A$ given as in
equation \eqref{gaugefield} takes the form
\be\label{axialcoordinates}
S_b(\phi, v, \vbar, A) = \frac{\kappa}{\pi}\int d^2 z (\p \phi + A_z )(\pbar\phi + A_{\zbar})
 + (\p_z +\p_z\phi+A_z)\vbar\, (\pbar+\pbar\phi + A_{\zbar})v \,.
\ee
A supersymmetric extension of the axially gauged action by worldsheet
fermions is provided by the addition of the fermionic action:
\be
S_f(\psi_{\pm},A) = \frac{\kappa}{\pi}\int d^2 z \big[ \psi^-(\p_{\zbar}+A_{\zbar})\psi^+ + \tilde{\psi}^-(\p_z+A_z)\tilde{\psi}^+  \big]\,.
\ee
The supersymmetry of the axial coset theory was analyzed in detail in \cite{Muto}.

\subsubsection{The $U(1)_R$ and a global $U(1)$ symmetry}

The axial coset model has $(2,2)$ supersymmetry.
In order to calculate the elliptic genus we must identify the
global $U(1)_R$ symmetry which is part of the left-moving $
N=2$ algebra. We follow the analysis in \cite{Henningson:1993nr}
in order to calculate the $U(1)_R$ charges of the fields.  Let us consider
a $U(1)$ transformation with parameter $ \gamma$ that acts on the
fermions as follows:
\begin{align}\label{Rsymm}
\psi^+ &\rightarrow e^{i\gamma c}\psi^+\qquad \psi^- \rightarrow
e^{-i\gamma c}\psi^-\cr \tilde{\psi}^+ &\rightarrow e^{-i\gamma
\tilde{c}}\tilde{\psi}^+\qquad \tilde{\psi}^- \rightarrow
e^{i\gamma \tilde{c}}\tilde{\psi}^- \,,
\end{align}
where $c$ and $\tilde{c}$ are arbitrary parameters for now. We
also postulate the following transformation for the bosonic fields
in the axially gauged WZW model:
\be
\delta g = i \gamma(\tilde{x} T g + x g T)\qquad\text{and}\qquad \delta A_{z,\zbar} = 0 \,.
\ee
Here $T$ is the Lie algebra valued matrix that is along the gauged direction. There are
three conditions imposed on the four variables $\{\tilde{c}, c, \tilde{x}, x\}$. The first follows from imposing that the action be invariant under the above transformations. The axially gauged bosonic action is classically non-invariant. Let us consider its variation, setting $x=0$ for now:
\be
\delta g = i\gamma \tilde{x} Tg \qquad \delta g^{-1}= -i\gamma \tilde{x} g^{-1 }T
\ee
The gauge field dependent terms vary as follows:
\begin{multline}
\frac{\pi}{\kappa}\delta {\cal L}_B = \Tr\bigg( A_{\zbar}\left[
-i\gamma \tilde{x} g^{-1}T\p_z g + g^{-1}\p_z(i\gamma \tilde{x}
Tg) \right] -A_z\left[\p_{\zbar}(i\gamma \tilde{x} Tg)g^{-1}
-\p_{\zbar}g (i\gamma \tilde{x} g^{-1}T)\right]\bigg)\cr
+\Tr\left[A_{\zbar}(-i\gamma \tilde{x} g^{-1}T)A_z g +
A_{\zbar}g^{-1}A_z (i\gamma \tilde{x} Tg)\right].
\end{multline}
Both the generator $T$ and the gauge fields $A_{z,\zbar}$ are
proportional to the same Lie algebra element, so they commute
with each other. Using this, one can see that the two terms in the
second line cancel each other out. Let us consider the term
proportional to the gauge field component $A_z$ in the first line.
Using integration by parts, we obtain for this term:
\be \Tr[\p_{\zbar}A_z (i\gamma \tilde{x} T) + A_z(i\gamma
\tilde{x} Tg)\p_{\zbar}g^{-1} + A_z(i\gamma \tilde{x}
\p_{\zbar}g)g^{-1} T]. \ee
Once again, commuting the generator $T$ through the gauge field
component $A_z$, the last two terms cancel. A similar calculation
can be done for the terms proportional to the gauge field
component $A_{\zbar}$. Putting them together, we find that
\be
\delta {\cal L}_B = (-i\gamma \tilde{x})\frac{\kappa}{\pi}\Tr(F_{z\zbar}T) \,.
\ee
A similar calculation can be done for the variation proportional
to $x$, leading to
\be
\delta S_B = \kappa\ \delta I_{A}(g,A) = i\gamma(x-\tilde{x})\frac{\kappa}{\pi}\int d^2z\ F_{z\zbar}\,.
\ee
The fermionic part is invariant classically but it has a quantum
anomaly due to the chiral anomaly in two dimensions. The anomalous
variation is:
\be
\delta S_F = 2i\gamma(\tilde{c}+c)\frac{1}{\pi}\int d^2z\ F_{z\zbar} \,.
\ee
Invariance of the action then imposes the constraint:
\be
\kappa(\tilde{x}-x) = 2(\tilde{c}+c)\,.
\ee
The remaining two constraints follow when we identify the ${\cal
N}=2$ supersymmetries in the axially gauged model and impose that
the  R-charge transformation commute with the right moving
supersymmetries and has unit charge under the left-moving
supersymmetries. This leads to the conditions:
\be
x = -c \qquad \text{and}\qquad \tilde{x} = \tilde{c} -1 \,.
\ee
The gauge symmetry of the action can  be used to impose the
constraint $x = -\tilde{x}$. Putting all this together, we find
the solution
\be c = \frac{1}{\kappa-2} =\frac{1}{k}\qquad\text{and}\qquad
\tilde{c}=\frac{\kappa-1}{\kappa-2} = \frac{k+1}{k}\,, \ee
where $k=\kappa-2$ is the supersymmetric level of the coset. These
equations determine the R-charges of the fermions, while the
charges of the bosons are given by
\be
x = -\frac{1}{k}\qquad\text{and}\qquad \tilde{x} = \frac{1}{k}\,.
\ee
%
This finalizes the determination of the left $U(1)_R$ action.

In addition, we have a global $U(1)$ symmetry that acts as a rotation on
the fermions and on the complex field $v$:
\begin{align}\label{globalsymm}
\psi^{\pm} \rightarrow e^{\mp i\lambda}\psi^{\pm} \qquad
\tilde{\psi}^{\pm} \rightarrow e^{\mp
i\lambda}\tilde{\psi}^{\pm}\qquad v\rightarrow e^{-i\lambda}v \,.
\end{align}
This is an isometric $U(1)$ rotation of the coset tangent space in
which these fields (or their derivatives) take values.

Our first goal is to compute the elliptic genus $ \chi_{cos}$,
twisted by the above global $U(1)$ symmetry:
\be \chi_{cos}(q,z,y) = \Tr\bigg((-1)^Fq^{L_0-\frac{c}{24}}\bar
q^{\bar L_0-\frac{c}{24}}\,  z^{J^R_0}\, y^{Q}\bigg) \,, \ee
where by $J^R_0$ and $Q$, we denote the zero-mode of the left R-current
and the global $U(1)$ charge respectively. We will also use the
notations $z=e^{2 \pi i \alpha}$ and $y=e^{2 \pi i \beta}$ for the chemical potentials. We
will refer to the above  quantity as the twisted elliptic genus.
We compute it in a Lagrangian picture using the path integral formalism. This involves
computing the partition function of the worldsheet theory on a
torus. The effect of the charge insertions in the twisted elliptic
genus will be to change the periodicity conditions of the charged
fields in the path integral. Modular covariance will be manifest at all stages.

\subsubsection{Parameterizing the gauge field with holonomies}

We will study the worldsheet theory on a torus, with the worldsheet coordinates identified under
the operations $(z,\zbar)\sim (z+2\pi, \zbar + 2\pi)\sim (z+2\pi \tau, \zbar+2\pi \bar\tau)$. The parameter $\tau = \tau_1 + i\tau_2$ is the modular parameter of the torus. To parameterize the group elements $h$ and $\bar{h}$ in terms of which we defined the gauge field, we introduce the function $\Phi$:
\begin{align}
\Phi(z,\zbar) &= \frac{i}{2\tau_2}\left[ s_1(z\bar{\tau}-\zbar
\tau) - s_2(\zbar-z)\right] \cr 
&=
\frac{i}{2\tau_2}(z\bar u-\zbar u)\,,
\end{align}
where we have defined $u=s_1\tau+s_2$. We take the holonomies $s_{1,2}$ to satisfy
$0\le s_1,
s_2 < 1$. The function $\Phi(z,\zbar)$ is a real harmonic function with
the following periodicity:
\be \Phi(z+2\pi, \zbar+2\pi) = \Phi(z,\zbar) + 2\pi s_1
\quad\text{and}\quad\Phi(z+2\pi \tau, \zbar+2\pi\bar\tau) =
\Phi(z,\zbar) - 2\pi s_2\,. \ee
%
%
%
There is an inherent ambiguity in the definition of the gauge field defined as in
equation  \eqref{gaugefieldsh}. The function $h$ can be multiplied by a purely anti-holomorphic function and it does not affect the expression for the gauge field $A_z$ and similarly, the field $\bar{h}$ can be multiplied by a holomorphic function without changing the gauge field $A_{\zbar}$. We make a particular choice to fix this ambiguity and parameterize the group elements $h$ and $\bar{h}$ which lead to the the gauge field in equation \eqref{gaugefieldsh} as follows:
\begin{align}\label{hhbar}
h(z,\zbar) &= e^{(X-iY)T}\, h^{u}\cr
\bar{h}(z,\zbar)&= (h(z,\zbar))^{\dagger} = (h^{u})^{\dagger}\, e^{(X+iY)T}\,.
\end{align}
where we have defined
\be\label{hudefn}
h^u =  e^{\frac{-T}{2\tau_2}\bar{u}(z-\zbar)} \quad\text{and}\quad (h^u)^{\dagger}=e^{\frac{T}{2\tau_2}u(z-\zbar)} \,.
\ee
The generator $T$ is the generator of the $U(1)$ subgroup that is being
gauged. The scalar field $X$ corresponds to a non-compact direction
while $Y$ is a compact boson which has non-trivial windings around the
cycles of the torus. The group elements $\tilde{h}$
and $\tilde{\bar h}$ are obtained from equation \eqref{hhbar} by a sign
flip of  the generator $T$. With these definitions, the gauge fields take the form
\begin{align}
A_z &= \p X -i\p Y-\frac{\bar u}{2\tau_2}\cr
&= \p X -i\p Y^{u}\,.
\end{align}
Here we have defined
\be\label{defYu}
Y^u(z,\zbar) = Y(z,\zbar) + \Phi(z,\zbar)\,.
\ee
Similarly, the anti-holomorphic component of the gauge field becomes
\be
A_{\zbar} = \pbar X+ i\pbar Y - \frac{u}{2\tau_2} = \pbar X +i \pbar \overline{Y^u} \,.
\ee
%

\subsection{The Twisted Partition Function}

We compute the twisted elliptic genus in the path integral formalism. We will discuss in detail the precise periodicity conditions to be imposed
 in the next subsection. We denote the partition function as
\be\label{pi} \chi_{cos}(q,z,y) =\int_{\Sigma} d^2u \int [{\cal
D}g] \int [{\cal D}X{\cal D}Y]\int [{\cal D}\psi_{\pm} {\cal
D}\tilde{\psi}^{\pm}] \, e^{-\kappa I_A(g,
A)}e^{-S_{f}(\psi^{\pm}, \tilde{\psi}^{\pm}, A)}\bigg |_{TPC} \,,
\ee
where the subscript refers to the twisted periodicity conditions. The fermionic measure is defined so as to respect the axial gauging. The integral over the gauge field has been broken up into an ordinary integral over the holonomy $u$ and the integral over the scalar fields $X$ and $Y$. In what follows we will show that the path integral factorizes into Gaussian integrals.

\subsubsection{Breaking up the bosonic action}
We start with the bosonic piece of the axially gauged model. Using the Polyakov-Wiegmann identity, we can rewrite it as follows:
\be\label{SAfinal}
I_A(g, h, \bar h) = I(h^{-1}g\tilde{\bar h}) - I(h^{-1}\bar h) \,.
\ee
Substituting equation \eqref{hhbar} into the action, we get
\begin{align}
I_A(g,A) = I((h^u)^{-1}e^{(-X-iY)T}\, g\, e^{(-X+iY)T} ((h^u)^{\dagger})^{-1}) - I((h^{u})^{-1}e^{-2iY\, T}(h^{u})^{\dagger}) \,.
\end{align}
Let us perform a similarity transformation on $g$, with
\be
g \longrightarrow g' = e^{(-X-iY)T}\, g\, e^{(-X+iY)T} \,,
\ee
and add and subtract the term $I(h^u\cdot ((h^u)^{\dagger})^{-1})$:
\begin{eqnarray}
I_A(g,A) &=& \bigg(I((h^u)^{-1}\, g'\, ((h^u)^{\dagger})^{-1}) -I(h^u\cdot ((h^u)^{\dagger})^{-1})\bigg)
\nonumber \\
& & 
 -\bigg( I((h^{u})^{-1}e^{-2iY\, T}(h^u)^{\dagger})-I(h^u\cdot ((h^u)^{\dagger})^{-1})\bigg).
\end{eqnarray}
Analogously  to equation \eqref{SAfinal} one can also write the vector gauged actions as follows:
\be\label{SVfinal}
I_V(g,h, \bar h) = I(h^{-1}g \bar h) - I(h^{-1}\bar h) \,.
\ee
Equations \eqref{SAfinal} and \eqref{SVfinal} imply that:
\be
I_A(g,A) = I_V(g', h^u, ((h^u)^{\dagger})^{-1}) - I_A(e^{-2iY\, T}, h^u, ((h^u)^{\dagger})^{-1}) \,.
\ee
The key point that makes this identity useful in doing
the path integral is the invariance of the measure for $g$ as a result of which one can replace ${\cal D}g$ with ${\cal D}g'$. The path integral therefore takes the form
\begin{multline}\label{piinitial}
\chi_{cos} =\int_{\Sigma} d^2u \int [{\cal D}g'] e^{-\kappa
I_V(g', h^u, ((h^u)^{\dagger})^{-1})}\cr \times \int [{\cal D}X]
[{\cal D}Y] e^{\kappa I_A(e^{-2iY\, T}, h^u,
((h^u)^{\dagger})^{-1})} \times \int [{\cal
D}\psi_{\pm}]e^{-S_{f}(\psi^{\pm}, \tilde{\psi}^{\pm}, A)}
\bigg|_{TPC}\,.
\end{multline}
The action of the abelian compact boson $Y$ is easy to evaluate:
\begin{align}
I_A(e^{-2iY\, T}, h^u, ((h^u)^{\dagger})^{-1}) &= -\frac{1}{\pi}\int d^2 z \left|i\p Y +\frac{\bar u}{2\tau_2}\right|^2 \cr
= -\frac{1}{\pi}\int d^2 z |\p\, Y^u|^2\,,
\end{align}
where we used equation \eqref{defYu}.

\subsubsection{Gauge degrees of freedom}

The axial coset model has a gauge symmetry. The non-compact $X$ field introduced above is precisely the gauge degree of freedom associated to this gauge symmetry, and therefore, it can be gauged away without affecting the physics. Gauge fixing the path integral leads to the addition of a $(b,c)$ ghost system via the Fadeev-Popov procedure. The ${\cal D}X$ integral therefore is just the volume of the gauge group. Dividing by this volume, we end up with the path integral
\begin{align}
\chi_{cos} &=\int_{\Sigma} d^2u \int [{\cal D}g] e^{-\kappa I_V(g,
h^u, ((h^u)^{\dagger})^{-1})} \times \int [{\cal D}Y] e^{\kappa
I_A(e^{-2iY\, T}, h^u, ((h^u)^{\dagger})^{-1})}\cr &\hspace{1.5in}
\times \int [{\cal D}\psi_{\pm}]e^{-S_{f}(\psi^{\pm},
\tilde{\psi}^{\pm}, A)} \times \int [{\cal D}b {\cal D}c {\cal
D}\tilde{b}{\cal D} \tilde{c}]
e^{-S_{gh}(b,c,\tilde{b},\tilde{c})}\bigg|_{TPC}\,.
\end{align}
At this point, the Euclidean $AdS_3$ action is decoupled from the
remaining fields. The fermions are still coupled to the $Y$-field but
we will disentangle these two sectors. We will write down the fully
factorized partition function after we discuss the twisted periodicity
conditions since both the twisting and the holonomies play a role in
decoupling the fermions.
\subsubsection{Twisted periodicity conditions}

We now turn to describe the periodicity conditions that we impose on
our fields.  Since we would like to put the $U(1)_R$ twist as an
operator insertion, this implies that the twist is in the time
direction. 
Since the Hamiltonian is $L_0$, the
definition of the trace singles out $\tau$ as the time direction and we
therefore put the twisted periodicity condition along the $\tau$-direction. {From}
the R-charges and global symmetry charges of the fields we get:
\begin{align}\label{TPC}
v(z+\tau, \zbar +\bar\tau) &= e^{i(\frac{\a}{k} -\b)}v(z,\zbar) \cr
\psi^{\pm} ( z + \tau, \bar z + \bar \tau) &= e^{\pm i (\frac{\a}{k}-\b)} \psi^{\pm}(z,\bar z)\cr
\tilde{\psi}^{\mp}  ( z + \tau, \bar z + \bar \tau) &= e^{\pm i (\frac{\alpha(k+1)}{k}-\b)} \tilde{\psi}^{\mp} (z,\bar z)\, .
\end{align}
Since it is  more straightforward to do the path integral over periodic fields,
we redefine the bosonic field $v$ such that it becomes periodic. We define
a new periodic field $v_p$:
\begin{align}\label{rotns}
v_p (z,\bar{z}) &= v(z,\bar{z}) e^{- (\frac{\alpha}{k}-\b) (z-\bar{z})/2 \tau_2}\,.
\end{align}
The effect of this on the action is clear. It modifies the holonomy
coupled to the field $v$ additively. There will be a similar effect on
the fermions but it is  more subtle since the $\a$-dependent
periodicity conditions can only be removed by an anomalous rotation of
the fermions. We describe this in detail next.

\subsubsection{The Fermionic Action}

The fermionic action is of the form
\be
S_f(\psi_{\pm},A) = \frac{\kappa}{\pi}\int d^2 z \big[ \psi^-(\p_{\zbar}+A_{\zbar})\psi^+ + \tilde{\psi}^-(\p_z+A_z)\tilde{\psi}^+ \big]\,.
\ee
Let us first perform a chiral rotation on the fermions and define new fields
\begin{align}\label{chiralrot}
\eta^{\pm} &= e^{\pm iY \pm \frac{u(z-\zbar)}{2\tau_2}} \psi^{\pm} \cr
\tilde{\eta}^{\mp} &= e^{\pm iY \pm \frac{\bar u(z-\zbar)}{2\tau_2}} \tilde{\psi}^{\mp}
\end{align}
The action now takes the form
\be
S_f =\frac{\kappa}{\pi} \int d^2z\ \big(\eta^+\p_{\zbar} \eta^- + \tilde{\eta}^+\p_z\tilde{\eta}\big) \,.
\ee
The new fermions $\eta$ and $\tilde{\eta}$ satisfy the periodicity conditions
\begin{align}\label{newbdry}
\eta^{\pm}(z+2\pi \tau, \zbar+2\pi\tau) &= e^{\pm i\left(u +\frac{\a}{k}-\b\right)}\eta^\pm(z,\zbar)\cr
\tilde{\eta}^{\mp}(z+2\pi \tau, \zbar+2\pi\bar\tau) &= e^{\pm i\left(\bar u +\frac{\a(k+1)}{k}-\b\right)}\tilde{\eta}^{\mp}(z,\zbar)
\end{align}
The chiral rotation we performed in equation \eqref{chiralrot} is
anomalous. This means that the fermionic measure transforms as
well.
If the fermions were periodic to begin with, the anomaly due to
the chiral rotation in equation \eqref{chiralrot} is given by
\be\label{yaddition}
-\frac{2}{\pi} \int d^2z\ \left |\p Y^u\right|^2 \,.
\ee
There is an additional contribution to the anomaly due to the
twisted periodicity condition on the fermions which is a pure
phase equal to:
\be
\int d^2z\frac{\a}{2\tau_2}(z-\zbar) F_{z\zbar}\,. \ee
Note that the $\b$-dependence of the
boundary conditions can be removed by a non-anomalous axial
rotation of the fermions. After integrating the anomalous phase by
parts,  it can be written as the wedge product of two one-forms:
\be
-i\int \frac{\a}{2\tau_2}(dz-d\zbar)\wedge dY^u \,.
\ee
Using the Riemann bilinear identity, we find that this is equal to
\be\label{phase}
2\pi \alpha (w+s_1) \,.
\ee
Therefore the net effect of the chiral rotation that gave rise to
the free fermion action is two-fold. Firstly, we get a bosonic
contribution to the action, whose result is to shift the
coefficient of the $Y$-part of the action, $\kappa\rightarrow
\kappa -2$. Secondly,  we get an $\a$-dependent phase as shown in
equation \eqref{phase}. Finally, we end up with a completely
factorized form of the partition function:
\begin{align}
\label{pifinal}
\chi_{cos} &=\int_{\Sigma} d^2u \int [{\cal D}g] e^{-\kappa I_V(g,
h^u, ((h^u)^{\dagger})^{-1})} \times \int [{\cal D}Y]
e^{(\kappa-2) I_A(e^{-2iY\, T}, h^u, ((h^u)^{\dagger})^{-1})}\cr
&\hspace{.1in} \times \int [{\cal D}\eta_{\pm} {\cal
D}\tilde{\eta}^{\pm}] e^{2 \pi i \alpha (w+ s_1)}\ e^{-S_{f}(\eta^{\pm},
\tilde{\eta}^{\pm})} \times \int [{\cal D}b {\cal D}c {\cal
D}\tilde{b}{\cal D} \tilde{c}]
e^{-S_{gh}(b,c,\tilde{b},\tilde{c})}\cr &= \int_{\Sigma} d^2u\
Z_g(u, \tau)\, Z_Y(u, \tau)\, Z_f(u, \tau)\, Z_{gh}(\tau)\,.
\end{align}
Note that the four pieces have no common factor such that the path
integrals can be performed separately.

\subsubsection{Evaluating the Partition Functions}

\noindent \underline{\bf The $H_3^+$ sector}: The vector-gauged
action in equation \eqref{piinitial} can be obtained by relating
it to an axially gauged action as follows:
\begin{align}
I_V(g, h^u, ((h^u)^{\dagger})^{-1}) &= I((h^u)^{-1}g ((h^u)^{\dagger})^{-1}) - I((h^u)^{-1}((h^u)^{\dagger})^{-1}) \cr
&= I_A(g, h^u, (h^u)^{\dagger}) + I((h^u)^{-1}(h^{u})^{\dagger}) - I((h^u)^{-1}((h^u)^{\dagger})^{-1})\,.
\end{align}
 The last two terms are easy to evaluate:
 \be
I((h^u)^{-1} (h^u)^{\dagger}) = \frac{-\pi (\Re u)^2}{\tau_2}\quad\text{and}\quad
I( (h^u)^{-1} ((h^u)^{-1})^{\dagger}) = \frac{\pi (\Im u)^2}{\tau_2}\,.
\ee
Substituting the values, we find that
\be
I_V(g, h^u, ((h^u)^{\dagger})^{-1}) =I_A(g, h^u, (h^u)^{\dagger}) -\frac{\pi |u|^2}{\tau_2} \,.
\ee
We have already written out the action for the axially gauged
coset in equation \eqref{axialcoordinates}. In this case the gauge
fields are purely given in terms of the holonomy. We find the
 action:
\begin{multline}
I_g = \kappa\, I_V(g, h^u, ((h^u)^{\dagger})^{-1}) = \frac{\kappa}{\pi}\int d^2 z (\p \phi -\frac{\bar u}{2\tau_2} )(\pbar\phi -\frac{u}{2\tau_2}) \cr
+ \frac{\kappa}{\pi}\int d^2 z (\p_z +\p_z\phi-\frac{\bar u}{2\tau_2})\vbar\, (\pbar+\pbar\phi -\frac{u}{2\tau_2})v
- \frac{\kappa \pi |u|^2}{\tau_2} \,.
\end{multline}
We now perform the chiral rotation of the fields $(v, \vbar)$ as in
equation \eqref{rotns} so as to write the action in terms of
periodic fields $v_p$ and $\vbar_p$. This leads to:
\begin{multline}
I_g =  \frac{\kappa}{\pi}\int d^2 z (\p \phi -\frac{\bar u}{2\tau_2} )(\pbar\phi -\frac{u}{2\tau_2}) \cr
+\frac{\kappa}{\pi}\int d^2 z \left(\p +\p\phi-\frac{\bar u-(\frac{\alpha}{k}-\b)}{2\tau_2}\right)\vbar_p\, \left(\pbar+\pbar\phi - \frac{u -(\frac{\alpha}{k}-\b)}{2\tau_2}\right)v_p
- \frac{\kappa \pi |u|^2}{\tau_2} \,.
\end{multline}
The path integral for the axially gauged action has been computed
previously \cite{Gawedzki:1991yu}. We quote the result for the
partition function:
\begin{align}
Z_{g}(u, \tau) =  \frac{\sqrt{k} \kappa}{\sqrt{\tau_2}} \frac{e^{\frac{2\pi (\Im
u)^2}{\tau_2}}}{|\theta_{11}\big(\tau,
u-\frac{\alpha}{k}+\b\big)|^2}.
\end{align}

\noindent \underline{\bf The Boson $Y$}: We recall the action of
the $Y$-dependent piece; the only subtlety is the shift in the
coefficient of the action, from $\kappa \rightarrow \kappa-2$,
which came from the anomalous rotation of the fermions:
\begin{align}
S_Y = -(\kappa-2) I_A(e^{-2iY\, T}, h^u, ((h^u)^{\dagger})^{-1}) = \frac{k}{\pi}\int d^2 z |\p\, Y^u|^2\,.
\end{align}
This is the action for a real compact scalar. Because of the presence of the holonomy, there is a shift in the periodicity of $Y^u$ around the cycles of the torus:
\begin{align}
Y^u(z+2\pi, \zbar + 2\pi) &= Y^u(z,\zbar) + 2\pi (w+s_1)\cr
Y^u(z+2\pi \tau, \zbar + 2\pi\tau) &= Y^u(z,\zbar) - 2\pi
(m+s_2)\,.
\end{align}
From the action, we observe that we have a twisted compact boson at radius $\sqrt{2k}$. The partition function for such a boson is given by
\be Z_Y(u,\tau) =\frac{\sqrt{k}}{\sqrt{\tau_2}|\eta(\tau)|^2}\sum_{m,n\in
\IZ} e^{-\frac{\pi k}{\tau_2}|(w+s_1)\tau +(m+s_2)|^2}. \ee

\noindent \underline{\bf The Fermions}: 
The action for the
fermionic part is of the form
\be
S_f(\eta^{\pm},\tilde{\eta}^{\pm}, a)  = \frac{\kappa}{\pi}\int d^2 z \left[\eta^+\p_{\zbar}\eta^- +  \tilde{\eta}^+\p_{z}\tilde{\eta}^- \right]\,.
\ee
The fermions $\eta$ and $\tilde{\eta}$ satisfy the periodicity
conditions specified in equations \eqref{newbdry}. The path integral for
such chiral fermions has been discussed for instance in
\cite{AlvarezGaume:1986es} and is given by\footnote{We have chosen
the phase factor for the chiral determinant such that the result
is covariant under modular S-transformations.}:
\begin{align}
  Z_f(u, \tau) &= \frac{1}{\kappa} \left[e^{-i2\pi s_1(s_2+\frac{\a(k+1)}{k}-\b)}
  e^{-2\pi\frac{(\Im u)^2}{2\tau_2}}
  \frac{\theta_{11}(\tau, u-\frac{\a(k+1)} {k}+\b)}{\eta(\tau)}\right]\times
  \\ & \quad \,
  \left[e^{i2\pi s_1(s_2+\frac{\a}{k}-\b)}e^{-2\pi\frac{(\Im u)^2}{2\tau_2}}
    \frac{\theta_{11}(\bar\tau, u-\frac{\a} {k}+\b)}{\eta(\bar\tau)} \right]
  \cr &= \frac{1}{\kappa} e^{-i2\pi s_1\alpha} e^{-2\pi\frac{(\Im u)^2}{\tau_2}}
  \frac{\theta_{11}(\tau, u-\frac{\a(k+1)}{k}+\b)\theta_{11}(\bar\tau, u-\frac{\a} {k}+\b)}{|\eta(\tau)|^2}\,.
\end{align}

\noindent \underline{\bf The Ghosts}: The ghost path integral is
standard:
\begin{align}
Z_{gh}(\tau) =  \int [{\cal D}b {\cal D}c {\cal D}\tilde{b}{\cal
D} \tilde{c}] e^{-S_{gh}(b,c,\tilde{b},\tilde{c})}=\ \tau_2\,
|\eta(\tau)|^4.
\end{align}
Putting all this together, we find that the full partition
function is:
\be \chi_{cos}(\tau, \a, \b) = k \int_0^1 d s_{1,2} \sum_{m,w \in
\mathbb{Z}} \frac{\theta_{11}(s_1 \tau + s_2 - \alpha
\frac{k+1}{k}+\b, \tau)}{\theta_{11} (s_1 \tau + s_2 -
\frac{\alpha}{k}+\b, \tau)} e^{2 \pi i \alpha w } e^{- \frac{ k
\pi}{\tau_2} |(m+s_2) + (w+s_1) \tau|^2}. \ee

\section{The long and short of it}
\label{contributions}


There are short multiplet or discrete character contributions to
the elliptic genus \cite{Eguchi:2004yi}, as well as long multiplet or continuous
character contributions \cite{Troost:2010ud}. In this section, we identify these two
types of contributions to the path integral. The result of the
axial coset path integral calculation was:
\begin{eqnarray}
\chi_{cos} &=& k \int_0^1 d s_{1,2} \sum_{m,w \in \mathbb{Z}}
\frac{\theta_{11}(s_1 \tau + s_2 - \alpha
\frac{k+1}{k}+\beta,\tau)}{\theta_{11} (s_1 \tau + s_2 -
\frac{\alpha}{k}+\beta,\tau)} e^{2 \pi i \alpha w } e^{- \frac{ k
\pi}{\tau_2} |(m+s_2) + (w+s_1) \tau|^2}.
\label{pathintegralresult}
\end{eqnarray}
Recall that we have the notations $z=e^{2 \pi i \alpha}$ as well as $y=e^{2 \pi
  i \beta}$.  To analyze the modular properties of the path integral
result, it is convenient to work with the expression after double
Poisson resummation:
\begin{eqnarray}
\chi_{cos} &=&  \int_0^1 d s_{1,2} \sum_{m,w \in \mathbb{Z}}
\frac{\theta_{11}(s_1 \tau + s_2 - \alpha \frac{k+1}{k}+\beta,\tau)}{\theta_{11} (s_1 \tau + s_2 - \frac{\alpha}{k}+\beta,\tau)}
e^{ - 2 \pi i s_2 w} e^{2 \pi i s_1 (m-\alpha)} e^{-  \frac{  \pi}{k \tau_2} |m-\alpha + w \tau|^2}.
\end{eqnarray}
The modular and elliptic properties can be computed as in \cite{Troost:2010ud}.
We summarize the result:
\begin{eqnarray}
\chi_{cos} (\tau+1,\alpha,\beta) &=& \chi_{cos} (\tau,\alpha,\beta) \nonumber \\
\chi_{cos} (-\frac{1}{\tau},\frac{\alpha}{\tau},\frac{\beta}{\tau}) 
&=& e^{ \pi i \frac{c}{3} \alpha^2 / \tau - 2 \pi i \alpha \beta / \tau} \chi_{cos} (\tau,\alpha,\beta) \nonumber \\
\chi_{cos} (\tau,\alpha+k,\beta) &=& (-1)^{\frac{c}{3} k} \chi_{cos} (\tau,\alpha,\beta) \nonumber \\
\chi_{cos} (\tau,\alpha+k \tau,\beta) &=& (-1)^{\frac{c}{3} k} e^{-  \pi i \frac{c}{3} (k^2 \tau + 2 k \alpha)}
e^{2 \pi i \beta k} \chi_{cos} (\tau,\alpha,\beta) \nonumber \\
\chi_{cos} (\tau,\alpha,\beta+1) &=& \chi_{cos} (\tau,\alpha,\beta) \nonumber \\
\chi_{cos} (\tau,\alpha,\beta+\tau) &=& e^{2 \pi i \alpha} \chi_{cos} (\tau,\alpha,\beta).
\end{eqnarray}
This is a Jacobi form in three variables, of weight zero, and with indices given by the above transformation
rules.

To analyze the
various parts of the spectrum that contribute to the path integral result, we Poisson resum on
the integer $m$ only in equation (\ref{pathintegralresult}) to go
to a Hamiltonian picture:
 \begin{eqnarray}
\chi_{cos}
 &=&    \sqrt{k \tau_2}  \sum_{n,w}
 \int \int ds_1 ds_2
  \frac{\theta_{11}(\tau,s_2+s_1 \tau-\alpha \frac{k+1}{k} +\beta)}{ \theta_{11} (\tau,s_2 + s_1 \tau -  \frac{\alpha}{k} + \beta)} e^{ 2 \pi i \alpha w}
 q^{\frac{(n - k(w+ s_1))^2}{4k}}
   \bar{q}^{\frac{(n+k (w+ s_1))^2}{4k}}
 e^{-2 \pi i s_2 n}. \nonumber
 \end{eqnarray}
 The details of the intermediate steps follow \cite{Troost:2010ud}
 closely, and we will therefore be brief.  We expand the
 theta-function in denominator and numerator, relabel summation
 variables, introduce the integral over the radial momentum $s$, and
 perform the integration over the holonomies $s_{1,2}$ to find:
 \begin{eqnarray}
  \chi_{cos}
&=&    \frac{1}{\pi}  \frac{1}{\eta^3}  \sum_{m,v,w}
\int_{-\infty-i \epsilon}^{+\infty - i \epsilon}
\frac{ds}{2is + v} (q^{is + \frac{v}{2}} \bar{q}^{is+\frac{v}{2}} -1)
\nonumber \\
& &
(-1)^m q^{\frac{(m-\frac{1}{2})^2}{2}}
 S_{v+m-kw-1}
q^{- v w + kw^2} z^{m-1/2} z^{-\frac{v}{k}+2 w}
 (q \bar{q})^{\frac{s^2}{k}+\frac{v^2}{4k}}y^{v-kw},
\end{eqnarray}
where  the special function $S_r(q)$ is defined by the formula:
\begin{eqnarray}
S_r (q) &=& \sum_{n=0}^{+ \infty} (-1)^n q^{\frac{n(n+2r+1)}{2}}.
\end{eqnarray}
As in \cite{Troost:2010ud}, we split this result into a holomorphic piece
and a remainder term. The holomorphic piece is equal to:
\begin{eqnarray}
\chi_{cos,hol}   &=&    \frac{1}{\pi}  \frac{1}{\eta^3}  \sum_{m,v,w}
\int_{-\infty-i \epsilon}^{+\infty - i \epsilon}
\frac{ds}{2is + v}(1-  S_{v+m-kw-1}+  S_{v+m-kw-1}q^{is + \frac{v}{2}} \bar{q}^{is+\frac{v}{2}}) \nonumber \\
& &
(-1)^m q^{\frac{(m-\frac{1}{2})^2}{2}}
q^{- v w + kw^2} z^{m-1/2} z^{-\frac{v}{k}+2 w} y^{v-kw}\  (q \bar{q})^{\frac{s^2}{k}+\frac{v^2}{4k}}\,,
\end{eqnarray}
which we can massage, using
the properties $q^r S_r = S_{-r}$ and
 $S_r + S_{-r-1} =1$ for the special function $S_r$, into a contour integral:
\begin{eqnarray}
\chi_{cos,hol}   &=&    \frac{1}{\pi}  \frac{1}{\eta^3}  \sum_{m,v,w}
\left( \int_{-\infty-i \epsilon}^{+\infty - i \epsilon}-\int_{-\infty-i \epsilon+i\frac{k}{2}}^{+\infty - i \epsilon+i\frac{k}{2}} \right)
\frac{ds}{2is + v} S_{-v-m+kw}(-1)^m q^{\frac{(m-\frac{1}{2})^2}{2}}
q^{- v w + kw^2} z^{m-1/2} z^{-\frac{v}{k}+ 2 w}
 \nonumber \\
& & y^{v-kw}
(q \bar{q})^{\frac{s^2}{k}+\frac{v^2}{4k}}.
\end{eqnarray}
The contour integral is easily performed. We pick up poles when the
radial momentum is equal to the angular momentum
$2is+v = 0$ for
values $2is$ in the interval
$0$ to $-(k-1)$. We therefore find the discrete
character contributions:
\begin{eqnarray}
\chi_{cos,hol}   &=&    \sum_{\gamma=0}^{k-1} \frac{1}{\eta^3}
\sum_{m,w}
S_{-\gamma-m+kw}(-1)^m q^{\frac{(m-\frac{1}{2})^2}{2}}
q^{-\gamma w + kw^2} z^{m-1/2} z^{-\frac{\gamma}{k}+ 2 w}
y^{\gamma-kw},
\end{eqnarray}
which  is equal to:
\begin{eqnarray}
\chi_{cos,hol}
&=& \sum_{\gamma \in \{0, \dots, k-1 \}} \sum_w
\frac{i \theta_{11}(\tau, \a)}{\eta^3}
\frac{q^{k w^2} q^{-w \gamma} z^{2w- \frac{\gamma}{k}}}{1-z q^{kw-\gamma}}
 y^{\gamma-kw}
\label{holomorphicpiece} \\
&=& \frac{1}{k} \sum_{\gamma,\delta \in \mathbb{Z}_k}
e^{\frac{2 \pi i \gamma \delta}{k}}
   \frac{i \theta_{11}(\tau, \a)}{\eta^3}
    \sum_{w \in \mathbb{Z}} \frac{
q^{\frac{(kw+\gamma)^2}{k}} z^{2 \frac{kw+ \gamma}{k}}    }{1-z^{\frac{1}{k}}
q^{w+ \frac{\gamma}{k}} e^{\frac{2 \pi i \delta}{k}} } y^{-(\gamma+kw)},\label{doublesum}
\end{eqnarray}
where we made the periodicity in the variable $\gamma$ manifest
in the last line.
The non-holomorphic remainder term can be rewritten as:
\begin{eqnarray}
\chi_{cos,rem} &=&
   -\frac{1}{\pi \eta^3}  \sum_{m,n,w}
\int_{-\infty-i \epsilon}^{+\infty - i \epsilon}
\frac{(-1)^mds}{2is + n+kw}  q^{\frac{(m-1/2)^2}{2}} z^{m-\frac{1}{2}}
y^n
q^{\frac{s^2}{k}+\frac{(n-kw)^2}{4k}} z^{\frac{kw-n}{k}}
\bar{q}^{\frac{s^2}{k}+\frac{(n+kw)^2}{4k}}.
\end{eqnarray}
We see that asymptotically, the global $U(1)$ symmetry has the
interpretation of measuring angular momentum on the cigar coset.

It can straightforwardly be checked that the full path integral
result for the twisted axial coset elliptic genus can be written
in terms of generalized Appell functions as follows:
\begin{eqnarray}
\chi_{cos} (q,z,y) &=&\frac{1}{k} \frac{i \theta_{11}(\tau, \a)}{\eta^3}    \sum_{\gamma,\delta \in \mathbb{Z}_k} e^{\frac{2 \pi
i \gamma \delta}{k}}
   q^{\frac{\gamma^2}{k}} z^{ \frac{2 \gamma}{k}} y^{-\gamma} z^{-1} q^{-\gamma}
\hat{A}_{2k} (z^{\frac{1}{k}} q^{\frac{\gamma}{k}} e^{\frac{2 \pi i \delta}{k}}, q^{-k+2 \gamma} z^{2} y^{-k}   ;q)
\nonumber \\
&=& \frac{1}{k} \frac{i \theta_{11}(\tau, \a)}{\eta^3}
\sum_{\gamma,\delta \in \mathbb{Z}_k} e^{-\frac{2 \pi i \gamma
\delta}{k}} q^{-\frac{\gamma^2}{k}}  y^{-\gamma} \hat{A}_{2k}
(z^{\frac{1}{k}} q^{\frac{\gamma}{k}} e^{\frac{2 \pi i
\delta}{k}},  z^{2} y^{-k}   ;q).
\end{eqnarray}
These generalized Appell functions were defined and analyzed in
\cite{Zwegers}.  It was rigorously proven there that they are real
Jacobi forms in three variables \cite{Zwegers}. Dressed with the
theta-functions, eta-functions, and the prefactors, the modular transformation
properties of the generalized Appell functions match those of our path integral
result for the twisted elliptic genus.

\section{Orbifold, spectrum, and spectral asymmetry}
\label{remarks}
In the previous section, we identified discrete and continuous
character contributions to the path integral result, and matched both
of these onto the theory of mock modular forms and their modular
completion. In this section, we would like to look at the physical
interpretations of these expressions in a bit more detail.  We will
relate the model to the one discussed in \cite{Troost:2010ud} and
generalize the latter to include the global $U(1)$ twist. We give an
interpretation of the holomorphic part in terms of individual free
field contributions, and in terms of characters. We also remark on the
global $U(1)$ charge as well as on how to derive the spectral density
of the non-holomorphic contributions via an independent method.

\subsection{Relation to its $\mathbb{Z}_k$ orbifold}
We wish to relate the previous result to the one obtained in
\cite{Troost:2010ud}. In order to do so, we can start with the result
we have above, and perform a $\mathbb{Z}_k$ orbifold, where
$\mathbb{Z}_k$ is a subgroup of the $U(1)_R$ symmetry. We perform this
orbifold as in \cite{Kawai:1993jk}, but the extra
$\beta$-dependence in the ellipticity properties of our twisted
elliptic genus leads to an extra $y-$dependence in the orbifold formula.
Another way to understand this dependence is by realizing that we introduce
twisted sectors by spectral flow. Spectral flow changes the boundary conditions
on the supercurrents, and therefore on the fermions. Since the fermions contribute
to the global $U(1)$ charge, twisting them also gives rise to an extra $y$-dependence
in the phase. Taking this into account, we obtain the expression:
\begin{eqnarray}
\chi_{orb,hol} &=& \frac{1}{k} \sum_{\tilde{\gamma},\tilde{\delta} \in \mathbb{Z}_k}
(-1)^{\tilde{\gamma} +\tilde{\delta}}
e^{2 \pi i \frac{\tilde{\gamma} \tilde{\delta}}{k}} q^{\frac{\tilde{\gamma}^2}{2}
+ \frac{\tilde{\gamma}^2}{k}}
z^{\tilde{\gamma} + \frac{2 \tilde{\gamma}}{k}} y^{- \tilde{\gamma}}
   \frac{i \theta_{11}(\tau, \alpha + \tilde{\gamma} \tau + \tilde{\delta})}{\eta^3}
   \nonumber \\
   & &
 \sum_{\gamma \in \{0, \dots, k-1 \} , m \in \mathbb{Z}}
\frac{q^{k m^2} q^{-m \gamma} z^{2m-\frac{\gamma}{k}}
e^{-2 \pi i \tilde{\delta} \frac{\gamma}{k}}
q^{\tilde{\gamma} (2m- \frac{\gamma}{k})}
}{1-z  q^{km-\gamma+\tilde{\gamma}}} y^{\gamma-km}
\nonumber \\
 &=&  \sum_{\gamma=0}^{k-1}
   \frac{i \theta_{11}(\tau, \alpha )}{\eta^3}
 \sum_{ m \in \mathbb{Z}}
\frac{q^{k m^2} q^{+m \gamma} z^{2m+\frac{\gamma}{k}}
}{1-z  q^{km}} y^{-km}
\nonumber \\
 &=&
   \frac{i \theta_{11}(\tau, \alpha )}{\eta^3}
 \sum_{ m \in \mathbb{Z}}
\frac{q^{k m^2}  z^{2m} y^{-km}
}{1-z^{\frac{1}{k}}  q^{m}}.
\end{eqnarray}
This is the coset conformal field theory $\mathbb{Z}_k$
orbifold whose path integral was computed in \cite{Troost:2010ud}.
Here, we have added a chemical potential coupling to an extra global
$U(1)$ symmetry. 
The non-holomorphic remainder term can also be computed by
orbifolding, or as in \cite{Troost:2010ud} directly from the path
integral result. We find:
\begin{eqnarray}
\chi_{orb,rem} &=&-\frac{1}{\pi}
\frac{1}{\eta^3}
\sum_{m \in \mathbb{Z}} (-1)^m q^{\frac{(m-\frac{1}{2})^2}{2}} z^{m-1/2}
\sum_{w \in \mathbb{Z}} \sum_{v \in \mathbb{Z} }
 y^{ k w} z^{\frac{v}{k}-2w}   q^{kw^2-vw}
\int_{-\infty-i \epsilon}^{+\infty-i \epsilon} \frac{ds}{2is+v} (q
\bar{q})^{\frac{s^2}{k}+\frac{v^2}{4k}}. \nonumber
\end{eqnarray}
Asymptotically, the global $U(1)$ charge corresponds to winding
number. The complete path integral result is:
\begin{eqnarray}
\chi_{orb} &=& \sum_{m,w} \int_{0}^{1} ds_1 ds_2 \frac{\theta_{11}
(\tau,  s_1 \tau + s_2- \frac{k+1}{k}\alpha+\beta )}{ \theta_{11}
(\tau, s_1 \tau + s_2-\frac{1}{k} \alpha+ \beta)} e^{\frac{2 \pi i
\alpha  w}{k}} e^{ - \frac{\pi}{k \tau_2} | (m+ks_2)+\tau (w+k
s_1)|^2}, \label{pathintegral2}
\\
&=&  \frac{i \theta_{11}(\tau,\alpha)}{\eta^3} \hat{A}_{2k} (z^{\frac{1}{k}},z^2 y^{-k} ; q),
\end{eqnarray}
which is the result of \cite{Troost:2010ud}, dressed with a twist.

If we were to apply the $\mathbb{Z}_k$ orbifold procedure once more, we would recuperate the
twisted axial coset partition function.  We note that the axial coset result
corresponds by T-duality \cite{Hori:2001ax}\cite{Israel:2004jt}
to $N=2$ Liouville theory at radius $R=
\sqrt{\alpha'/k}$ and exhibits a single ground state which is in
accord with the Witten index calculation of \cite{Girardello:1990sh}, which
states that the number of ground states is equal to the radius divided
by $\sqrt{\alpha'/k}$. 
The orbifolded coset has $k$ ground states as can be seen by putting the twists to zero,
$\alpha=0=\beta$. There are as many ground states as in $N=2$ Liouville theory at
radius $R=\sqrt{\alpha' k}$
\cite{Girardello:1990sh}\footnote{Reference \cite{Girardello:1990sh}
  regularizes $N=2$ Liouville theory such that there are $k$ ground
  states. One of them belongs to the family of delta-function
  normalizable states. See e.g. \cite{Ashok:2007ui} for a
  discussion. In our context, the delta-function normalizable ground state leads to a
  minor ambiguity in how to split the elliptic genus into a holomorphic part and a
  non-holomorphic remainder. This ambiguity is of little consequence.}.

\subsection{Interpretation}
In this subsection we comment on the contribution of individual states
and discuss the $N=2$ superconformal character content of the holomorphic part
of the elliptic genus.
\subsubsection*{Regularized individual contributions}
We can compare the interpretation of the elliptic genus computed here
to that given in \cite{Troost:2010ud}.  Indeed, the free field
interpretations of the orbifold result obtained in
\cite{Troost:2010ud} have counterparts for the axial coset. For
instance, under the assumptions that $|q|<|z| < 1$ and
$y=1$, we can expand the holomorphic part of the axial coset as
follows (starting from equation \eqref{doublesum})
\begin{multline}
\chi_{orb,hol} = \frac{1}{k} \sum_{\gamma,\delta \in \mathbb{Z}_k}
e^{2 \pi i \frac{\gamma \delta}{k}}  
   \frac{i \theta_{11}(\tau, \alpha )}{\eta^3} 
\left( \sum_{k m + \gamma \ge 0, p \ge 0}- \sum_{k m + \gamma < 0, p < 0}
\right)
q^{\frac{(km+\gamma)^2}{k}} z^{2 \frac{km+\gamma}{k}} 
z^{\frac{p}{k}} e^{2 \pi i \frac{p \delta}{k}} q^{\frac{p (km+\gamma)}{k}}.
\nonumber
\end{multline}
We can put $km + \gamma = -w$ and $p=w+kn$ to find:
\begin{eqnarray}
\chi_{orb,hol} &=& \frac{1}{k} \sum_{\gamma,\delta \in \mathbb{Z}_k}
e^{2 \pi i \frac{\gamma \delta}{k}}  
   \frac{i \theta_{11}(\tau, \alpha )}{\eta^3} 
\left( \sum_{w \le 0, w+kn \ge 0}- \sum_{w > 0, w+kn < 0}
\right)
z^{\frac{(-w+kn)}{k}} e^{2 \pi i \frac{w \delta}{k}} q^{-wn}
\nonumber \\
 &=& 
   \frac{i \theta_{11}(\tau, \alpha )}{\eta^3} 
\left( \sum_{w \le 0, w+kn \ge 0}- \sum_{w > 0, w+kn < 0}
\right)
z^{\frac{kn-w}{k}}  q^{-nw}
\end{eqnarray}
Clearly, this is very much like equation (14) in
\cite{Troost:2010ud}, with the important difference that we are at
the inverse radius. As a consequence, we are summing individual
contributions over different wedges of the (momentum,winding)
plane.
\subsubsection*{Discrete character sum}
The interpretation of the holomorphic part (at $y=1$) as a sum over discrete
characters on the $N=2$ superconformal algebra is as follows. We
consider Ramond ground states of spin $j$ and R-charge
$\frac{2j-1}{k}-\frac{1}{2}$ with $N=2$ superconformal character (see e.g. \cite{Israel:2004jt}):
\begin{eqnarray}
ch^{\tilde{R}}_d(j;\tau,\alpha) &=& z^{\frac{2j-1}{k}} \frac{1}{1-z} 
\frac{i \theta_{11}(\tau,\alpha)}{\eta^3},
\end{eqnarray}
and spectrally flow them $-(2j-1)$ units to obtain:
\begin{eqnarray}
ch(j;\tau,\alpha) &=& z^{\frac{1-2j}{k}} \frac{1}{1-z q^{1-2j}}  
\frac{i \theta_{11}(\tau,\alpha)}{\eta^3}.
\end{eqnarray} 
We recognize this as the $m=0$ contribution to the holomorphic part of
the axial coset elliptic genus in equation (\ref{holomorphicpiece})
when summed over the spins $2j-1=0,\dots,k-1$. The full holomorphic
contribution is obtained by summing over the extension of these
characters by spectral flow by multiples of the level $k$.  That settles
the $N=2$ superconformal content of the holomorphic contribution.

\subsection{Spectral asymmetry}
Finally, we provide an independent way to derive the measure of the
non-holomorphic contribution to the elliptic genus. The origin of the
remainder contribution is a mismatch in the spectral density of
right-moving bosons and right-moving fermions.  If we concentrate on
right-movers only, the elliptic genus reduces to a Witten index. The
fact that we can obtain a contribution to the elliptic genus from a
continuum of modes due to a mismatch in the spectral density of boson
and fermions is known.

The relative spectral density between the two right-moving Ramond
sectors can be read off from the ratio of reflection amplitudes in
these sectors. The reflection amplitudes are given by
\cite{Israel:2004jt}:
\begin{eqnarray}
R^\pm (j,m \bar{m}) &=&
\frac{\Gamma(-2j+1)\Gamma( 1 + \frac{2j-1}{k}) \Gamma(j+m \mp 1/2) \Gamma( j- \bar{m} \pm 1/2)}{\Gamma(2j-1)\Gamma(1-\frac{2j-1}{k}) \Gamma(-j+1+m \mp 1/2)\Gamma(-j+1-\bar{m} \pm 1/2)},
\end{eqnarray}
where for continuous modes we put $j=\frac{1}{2} + i s$ where $s \in
\mathbb{R}$ and where $m, \bar{m}$ are the left- and right-moving
momentum respectively.  The spectral asymmetry (or difference in
spectral densities) in the two right-moving Ramond sectors is then
given by:
\begin{eqnarray}
\Delta \rho (s) = \rho_+(s) - \rho_-(s) &=& \frac{1}{2 \pi i} \frac{d}{ds}
\log \frac{R^+}{R^-}.
\end{eqnarray}
Using these formulae, and the fact that we change Ramond sectors
for the right-movers only, we obtain a spectral asymmetry:
\begin{eqnarray}
\Delta \rho (s) &=&
\frac{1}{2 \pi } \left( \frac{1}{is - \bar{m}} - \frac{1}{is + \bar{m}}
                 \right).
\end{eqnarray}
This is a spectral measure on the half-line $s \in {[} 0 , \infty {[}$.
If we integrate the measure against an even function of the radial momentum
$s$, the measure on the full line becomes
\begin{eqnarray}
-\frac{1}{2 \pi} \frac{1}{is + \bar{m}}
\end{eqnarray}
which agrees with the measure in the non-holomorphic remainder function, including
the normalization (after the appropriate identification $v = 2
\bar{m}$). Thus, that provides a direct justification of the
measure in the non-holomorphic contribution to the elliptic genus.
It gives a direct physical interpretation to the remainder
function of \cite{Zwegers}.

\section{Conclusions}
In this paper we have given a detailed path integral derivation of the
elliptic genus of a non-compact conformal field theory, further
twisted by a global $U(1)$ symmetry. We identified the short, discrete
contributions with a mock modular form, and the long, continuous
contributions as arising from a difference in spectral densities for
right-moving fermions and bosons. The whole result is a
(non-holomorphic) Jacobi form in three variables.  It is possible to
generate many further examples of (twisted) elliptic genera, and mock
modular forms related to Jacobi forms by orbifolding combinations of
non-compact and compact $N=2$ superconformal models. It will be
interesting to further investigate this class of forms. In particular,
they will have applications to checks on mirror symmetry for
non-compact Gepner and Landau-Ginzburg models (see
e.g. \cite{Eguchi:2000tc}\cite{Ashok:2007ui}), including the long
multiplet sector. It will also be interesting to attempt to apply
these ideas to prove duality properties of black hole entropy counting
functions from first principles.

\section*{Acknowledgements}
We would like to thank Luca Carlevaro, Atish Dabholkar, Jan
Manschot and Sameer Murthy for interesting discussions and
correspondence. S.A. would like to thank the Perimeter Institute
for hospitality during the completion of this work. The work of
J.T. is supported in part by the grant ANR-09-BLAN-0157-02.

\end{document}